\documentclass[10pt,conference,letterpaper]{IEEEtran}
\IEEEoverridecommandlockouts %
\usepackage{cite,amsmath,amssymb,amsfonts,graphicx,textcomp,xcolor,booktabs,url}

\newtheorem{proposition}{Proposition}
\makeatletter
\@ifundefined{pdfpagewidth}{}{\pdfpagewidth=8.5in \pdfpageheight=11in}
\@ifundefined{pagewidth}{}{\pagewidth=8.5in \pageheight=11in}
\makeatother

\usepackage{eso-pic}
\begin{document}
\AddToShipoutPictureFG*{\AtPageLowerLeft{\makebox[\paperwidth]{\raisebox{0.4in}{\parbox{\textwidth}{\footnotesize \copyright~2027 IEEE. Personal use of this material is permitted. Permission from IEEE must be obtained for all other uses, in any current or future media, including reprinting/republishing this material for advertising or promotional purposes, creating new collective works, for resale or redistribution to servers or lists, or reuse of any copyrighted component of this work in other works.}}}}}

\title{Curvature-Calibrated Friction and an Exact\\Gauge Symmetry for Quantum Hamiltonian Descent}
\author{\IEEEauthorblockN{Kentaro Hoshisashi}
\IEEEauthorblockA{\textit{Department of Computer Science, University College London}, London, United Kingdom\\
k.hoshisashi@ucl.ac.uk}}
\maketitle

\begin{abstract}
A design problem constrained by a partial differential equation can be solved coherently on a quantum computer. The objective is block-encoded rather than measured, and Quantum Hamiltonian Descent consumes it directly.
We identify two obstacles in this pipeline and repair both.
First, the oracle construction assumes affine generators with positive semidefinite Hermitian parts, which covariance entries violate for every off-diagonal generator. Encoding the covariance as an affine combination of positive semidefinite atoms restores both conditions at every register point, at the price of a representational floor, median $45\%$ of the fitted family's own objective range for a fixed rank-three dictionary and $22\%$ with a variable diagonal.
Second, adding a constant to the objective leaves the measurement distribution exactly invariant, yet the mean-objective friction heuristic in current use assigns such objectives different schedules. Calibrating the friction to the landscape curvature restores the symmetry, though not the constant in $\nu=c\sqrt\mu$, and a bracketed sweep at matched objective-phase exposure, not end-to-end cost, places the best tested friction between $0.032$ and $0.25$ of critical damping on a diffusion-design benchmark, a broad plateau an order of magnitude below the classical default.
\end{abstract}


\section{Introduction}
\label{sec:intro}

\begin{figure*}[!t]
\centering
\includegraphics[width=0.8\textwidth]{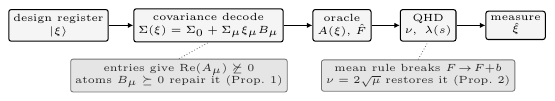}
\caption{The coherent design pipeline of Sec.~\ref{sec:intro}.}
\label{fig:pipeline}
\end{figure*}

Designing under a partial differential equation means choosing the parameters of an evolution equation so that its solution, at a fixed horizon, does what one wants.
Write the design variables as $\xi$ ranging over a box $\Xi$ and the problem is
\begin{equation}
\min_{\xi\in\Xi}\ F(\xi),\qquad
F(\xi)=\big\|P\big(u(T;\xi)-u^\star\big)\big\|^{2},
\label{eq:problem}
\end{equation}
in which $u$ evolves under $\dot u=-A(\xi)u$ from a fixed initial state and $P$ projects onto whatever part of it the objective observes.
Calibration, inverse design and optimal material layout all take this form.
The classical route is the adjoint method \cite{plessix2006adjoint}, whose cost is set by the need to hold or recompute the forward trajectory \cite{griewank2000algorithm}.
A recent line of work runs \eqref{eq:problem} coherently on a quantum computer \cite{sato2025blockencoding}.
The objective is never measured at an iterate.
It is block-encoded as an operator $\hat F$ on a register holding the design variables, and a continuous-time quantum optimiser, Quantum Hamiltonian Descent \cite{leng2023qhd}, consumes it directly (Fig.~\ref{fig:pipeline}).
The design register explores $\Xi$ in superposition, so no classical per-design forward trajectory is ever materialised or read out.

\begin{figure}[!t]
\centering
\includegraphics[width=\columnwidth]{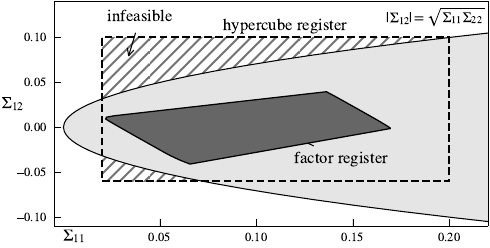}
\caption{Register, constraint and factor family in one slice. The pale lens is the positive semidefinite cone, the dashed rectangle the design box, the hatched corners its infeasible part, and the dark polygon the image of factor form \eqref{eq:factor}.}
\label{fig:feasible}
\end{figure}

Analysis of these pipelines has concentrated on the oracle, on how to block-encode $A(\xi)$ \cite{gilyen2019qsvt,low2019qubitization}, how to solve the resulting linear systems \cite{hhl2009}, how to simulate non-unitary dynamics \cite{an2023lchs,berry2014ode}, and what the resulting gate counts come to.
Resource analyses of digital QHD likewise treat the encoding of the search space as a first-class cost \cite{liu2026encoding}.
This paper works one level up, on what the design register may contain and on the schedule that traverses it.
Everything reported below is a property of the design landscape and of the schedule. What does depend on the construction of \cite{sato2025blockencoding} is the generator-wise assumption examined in Sec.~\ref{sec:feasible}, the prefactor in \eqref{eq:exposure} and the design-space discretisation, whose weight Sec.~\ref{sec:disc} measures.

At that level the pipeline meets two obstacles. The oracle construction assumes $A(\xi)=A_0+\sum_\mu \xi_\mu A_\mu$ affine with $\mathrm{Re}(A_\mu)\succeq0$ for every generator separately, and covariance entries fail the second condition outright, the generator of an off-diagonal entry carrying the sign-indefinite symbol $k_ik_j$. And the schedule rule in current use breaks an exact symmetry, since adding a constant to the objective changes the evolution by a global phase alone and leaves the measurement distribution exactly invariant, yet the mean-objective rule assigns gauge-equivalent objectives different schedules.

Both admit simple repairs. An affine combination of positive semidefinite atoms satisfies the oracle's conditions at every register point, and reading the friction from the curvature, which \cite{sato2025blockencoding} itself prefers where the curvature is known, restores the symmetry. Constraint handling is long-studied in evolutionary computation \cite{michalewicz1996constrained,coello2002survey,deb2000constraint,hansen2001cmaes}, and what is unusual here is that the requirement comes from the evaluation mechanism rather than from the search, ruling out the otherwise obvious Cholesky factorisation.

The contributions are three.
\begin{itemize}\setlength{\itemsep}{1pt}\setlength{\parskip}{0pt}\setlength{\topsep}{2pt}
\item A feasibility encoding for covariance-valued design, positive semidefinite atom by atom, with factor form as its rank-one case (Sec.~\ref{sec:feasible}).
\item A measurement of its price, a representational floor of median $45\%$ of the fitted family's own objective range, $22\%$ with a variable diagonal, and an order-of-magnitude conditioning spread (Secs.~\ref{sec:gap} and~\ref{sec:conditioning}).
\item The exact statement of the gauge symmetry and a bracketed sweep at fixed exposure, placing the best tested friction between $0.032$ and $0.25$ of critical damping across five draws (Secs.~\ref{sec:fricmeth} and~\ref{sec:sweep}).
\end{itemize}

Every measurement below was made classically, most in minutes on one CPU core and the two largest sweeps on one GPU, with the optimiser integrated as a Schr\"odinger equation on the design register alone, $M$ design variables of $m$ bits each, whose dimension $2^{mM}$ does not grow with the physical state. 

\section{Preliminaries}
\label{sec:background}

Everything below rests on one property of the oracle, that the objective is available as a diagonal operator $\hat F$ on an $M$-variable design register of $m$ bits each, and how that operator is built is treated here as settled.

QHD evolves a wavefunction $\psi(s)$ on the design register under
\begin{equation}
H(s) \;=\; -\frac{1}{2\lambda(s)}\,\Lambda^{(M)} \;+\; \lambda(s)\,\hat F ,
\qquad \lambda(s)=e^{\nu s},
\label{eq:qhd}
\end{equation}
with $\Lambda^{(M)}$ the design-space Laplacian and $\nu>0$ the friction rate.
Growing $\lambda$ transfers weight to the potential term and $|\psi(S)|^2$ concentrates on low values of $F$.
Hamilton's equations for the classical counterpart $H=p^2/2\lambda(s)+\lambda(s)F(x)$ give $\dot x=p/\lambda$ and $\dot p=-\lambda\nabla F$, so
\begin{equation}
\ddot x \;=\; \frac{\dot p}{\lambda}-\frac{p\,\dot\lambda}{\lambda^{2}}
\;=\; -\nabla F(x) - \nu\,\dot x ,
\label{eq:damped}
\end{equation}
the constant-friction flow underlying heavy-ball and Nesterov acceleration \cite{polyak1964,nesterov1983,su2016ode,wibisono2016variational}.

Two quantities recur below.
The first is the curvature of the landscape after the objective has been rescaled to the unit interval over the design box,
\begin{equation}
\tilde F(\xi)=\frac{F(\xi)-F_{\min}}{F_{\max}-F_{\min}},
\qquad
\mu=\lambda_{\min}\!\left(\nabla^2\tilde F(\xi^\star)\right),
\label{eq:rescale}
\end{equation}
with $L=\lambda_{\max}$ and $\kappa=L/\mu$, taken at the minimiser $\xi^\star$ in normalised design coordinates.
The second is the schedule integral
\begin{equation}
E \;=\; 6\,\alpha_{\mathrm{obj}} \int_0^{S}\lambda(s)\,ds
\;=\; 6\,\alpha_{\mathrm{obj}}\,\frac{e^{\nu S}-1}{\nu},
\label{eq:exposure}
\end{equation}
the leading objective-oracle contribution under the simulation accounting of \cite[App.~C]{sato2025blockencoding}. It is not an end-to-end query count, and it prices neither the kinetic term nor the classical estimation of $\mu$.
We refer to $E$ as the \emph{objective-phase exposure} and not as a query complexity.

Reference \cite{catli2025efficient} bounds the running time to an $\epsilon$-optimal solution by $S=\nu^{-1}\log(a/\epsilon)$, with $a$ set by the moments of the initial distribution and by $L$. Setting $a=1$, as the terminal rule $\lambda(S)=1/\epsilon$ does, discards that constant and with it the guarantee.
Where $\epsilon$ appears below it is the terminal-$\lambda$ parameter of the adopted rule.
Throughout, $\alpha_{\mathrm{obj}}=2.51$ and $\epsilon=4.5\times10^{-5}$, computed for the Black--Scholes example of \cite{sato2025blockencoding}.

\section{Method}
\label{sec:method}
 
The method has two parts, an encoding that keeps the oracle's assumptions valid across the whole register, and a friction rule that respects an exact symmetry of the dynamics.
 
\subsection{Feasible encoding}
\label{sec:feasible}
 
The oracle construction of \cite{sato2025blockencoding} rests on two structural assumptions about the generator.
The first is affinity in the design variables,
\begin{equation}
A(\xi)\;=\;A_0+\sum_{\mu=1}^{M}\xi_\mu A_\mu,
\label{eq:affine}
\end{equation}
which is what allows a single block encoding to serve every design point, the register selecting the coefficients coherently.
The second is that
\begin{equation}
\mathrm{Re}(A_\mu)\;\succeq\;0\qquad\text{for each $\mu$ separately},
\label{eq:genwise}
\end{equation}
in the Loewner order, a sufficient condition for $\mathrm{Re}(A(\xi))\succeq0$ over the whole box and the condition actually assumed in \cite{sato2025blockencoding}, whose Sec.~II-A states ``we assume that $\mathrm{Re}(A_\mu)$ is positive semidefinite for all $\mu$'' in order to guarantee the former.
The two conditions are not equivalent.
 
For a multi-asset diffusion the parameters in which the generator is affine are the entries of the covariance matrix $\Sigma$.
At wavenumber $k$ the symbol of $A$, its eigenvalue on that Fourier mode, reads
\begin{equation}
a(k,\xi)\;=\;r+\tfrac12\,k^{\mathsf T}\Sigma(\xi)\,k-i\,c(k,\xi),
\label{eq:symbol}
\end{equation}
with $r$ the short rate and $c$ collecting the drift terms, which do not enter the Hermitian part.
The pointwise condition $\mathrm{Re}\,a\ge r>0$ for every $k\in\mathbb{R}^d$ therefore holds exactly when $\Sigma(\xi)\succeq0$.
On a finite spectral grid positive semidefiniteness remains sufficient but not necessary, since a small negative eigenvalue $-\eta$ still leaves $\mathrm{Re}\,a>0$ on every sampled mode whenever $\tfrac12\eta K^{2}<r$, with $K$ the largest wavenumber represented.
 
The generator-wise condition \eqref{eq:genwise} is stronger, and differentiating \eqref{eq:symbol} shows where it breaks.
The generators attached to the entries of $\Sigma$ have symbols
\begin{equation}
\partial_{\Sigma_{ii}}\,\mathrm{Re}\,a=\tfrac12\,k_i^{2},
\qquad
\partial_{\Sigma_{ij}}\,\mathrm{Re}\,a=k_i k_j\quad(i\ne j).
\label{eq:gensym}
\end{equation}
The first is non-negative on every mode.
The second takes both signs across the mode set, so the generator of every off-diagonal entry violates \eqref{eq:genwise} outright, independently of the design box.
Pointwise feasibility fails too, though only in the box-dependent way of Fig.~\ref{fig:feasible}.
 
\subsection{A parameterisation}
 
Numerical optimisation has a standard answer here.
Encode a Cholesky factor $R$ and set $\Sigma=R^{\mathsf T}R$, positive semidefinite whatever $R$ is.
That answer makes $\Sigma$ quadratic in the design variables and so is incompatible with an affine-generator oracle, though not with constructions that replace the oracle layer.
What survives is a parameterisation that is affine and whose affine generators are themselves feasible.
 
\begin{proposition}
Consider a constant-coefficient diffusion generator whose Fourier symbol has Hermitian part $\tfrac12 k^{\mathsf T}\Sigma(\xi)k$, and let
\begin{equation}
\Sigma(\xi)\;=\;\Sigma_0+\sum_{\mu=1}^{M}\xi_\mu B_\mu,
\qquad \Sigma_0\succeq0,\quad B_\mu\succeq0.
\label{eq:atoms}
\end{equation}
Then $A(\xi)$ is affine in $\xi$, $\Sigma(\xi)\succeq0$ holds at every point of $[0,1]^{M}$, and each generator has $\mathrm{Re}(A_\mu)\succeq0$, its symbol being $\tfrac12\,k^{\mathsf T}B_\mu k\ge0$.
\end{proposition}
 
Every claim is read off \eqref{eq:atoms} directly.
Affinity is the form of the sum, pointwise feasibility follows because a non-negative combination of positive semidefinite matrices is positive semidefinite, and the generator-wise condition holds atom by atom.
The factor form
\begin{equation}
\Sigma(\xi)\;=\;\mathrm{diag}(\delta)+\sum_{\mu=1}^{M}\xi_\mu\,v_\mu v_\mu^{\mathsf T},
\qquad \delta_i\ge0,
\label{eq:factor}
\end{equation}
is the rank-one case, $B_\mu=v_\mu v_\mu^{\mathsf T}$, whose generator symbol is $\tfrac12(v_\mu\!\cdot\!k)^{2}$, and the condition $\delta_i\ge0$ is needed, since without it $\Sigma_0$ is itself infeasible.
 
Covariance entries need $m\,d(d+1)/2$ design qubits and \eqref{eq:factor} needs $mM$, which at fixed $M$ does not grow with $d$.
The loadings $v_\mu\in\mathbb{R}^{d}$, the coefficient oracle, the forward state register and the number of factors needed to approximate a given covariance all retain their $d$ dependence.
 
\subsection{Curvature-calibrated friction}
\label{sec:fricmeth}
 
\begin{proposition}
Let $F_b=F+b$ for a constant $b$. Then
\begin{equation}
H_b(s)=H(s)+b\,\lambda(s)\,I,
\quad
U_b(S)=e^{-i b\int_0^{S}\lambda(s)\,ds}\,U(S),
\label{eq:gauge}
\end{equation}
so $|\langle\xi|\psi_b(S)\rangle|^{2}=|\langle\xi|\psi(S)\rangle|^{2}$ for every state $\xi$ and every $b$.
\end{proposition}
 
The friction rule in current use reads the mean of the objective over the design box,
\begin{equation}
\nu_{\mathrm{heur}}\;=\;\sqrt{\overline F},
\label{eq:meanrule}
\end{equation}
which is not invariant, since $\overline{F_b}=\overline F+b$, so two descriptions of one problem are handed to two different algorithms, whose schedules differ, whose finite-time outputs need not agree, and whose selection depends on a convention rather than on the problem.
 
Normalising the objective by a norm bound $\gamma_u^{2}$, as \cite{sato2025blockencoding} does, fixes the multiplicative scale but leaves the additive freedom untouched.
Reading the friction from the curvature instead,
\begin{equation}
\nu\;=\;2\sqrt{\mu},
\label{eq:nurule}
\end{equation}
with $\mu$ the curvature of \eqref{eq:rescale}, is invariant under both the shift and the rescaling, and \cite{sato2025blockencoding} already recommends a curvature-based choice where the curvature is known.

\section{Experimental Setup}
\label{sec:setup}

\noindent \textbf{Benchmark.} The forward problem throughout is a constant-coefficient $d$-asset diffusion in log price on a periodic spectral grid of width $6$ with $N_x$ points per direction, rate $r=0.05$ and horizon $T=1$.
The claim $u_0$ is a difference of two Gaussians of width $0.8$ offset by $0.9$, normalised to unit length.
The objective is the full-domain squared residual against $u(T;\xi^\star)$ evaluated in the spectral domain by Parseval, so $P$ is the identity and $\mathrm{rank}(P)=N_x^{d}$.
Factor models draw $\delta_i$ uniformly from $[0.01,0.04]$ and the columns of $V$ from $\mathcal{N}(0,I/d)$ scaled by $0.55$.
Truncation to the $2\times10^{4}$ heaviest modes, which carry $99.88\%$ of the energy and move the rescaled landscape by $1.4\times10^{-3}$ without changing its minimiser, appears only in the spatial-refinement checks, and the $d=6$, $N_x=4$ landscapes carrying every other experiment use all $4096$ of their modes.

The minimiser $\xi^\star$ comes from Nelder--Mead, and $F_{\min}$ and $F_{\max}$ from a $13$-point-per-axis scan of the box.
For the in-family targets $F_{\min}=0$ by construction, and using that value instead shifts $\mu$ by under $0.1\%$. $F_{\max}$ agrees to nine figures at $13$, $17$ and $25$ points per axis, the maximum at a box corner.
The Hessian is a central difference with step $5\times10^{-3}$, its eigenvalues unchanged from step $2\times10^{-2}$ to $1.25\times10^{-3}$.
The design register carries $m$ bits per variable on an endpoint-inclusive grid of spacing $h_\xi=1/(2^m-1)$, and $\Lambda^{(M)}$ is built from
$\Lambda_{\mathrm{FD}}=h_\xi^{-2}\,\mathrm{tridiag}(1,-2,1)\preceq0$, the matrix of \cite[Eq.~(19)]{sato2025blockencoding}, without wrap-around.
The kinetic term of \eqref{eq:qhd} is therefore $-\Lambda^{(M)}/2\lambda\succeq0$, the sign the continuum operator $-\Delta/2\lambda$ requires.
The initial state is the uniform superposition and \eqref{eq:qhd} is integrated by Strang splitting at the step sizes stated in each table.

\label{sec:protocol}

\noindent \textbf{Protocol.} The entry parameterisation is sampled over the box with diagonal entries uniform in $[0.02,0.20]$ and off-diagonal entries in $[-0.06,0.10]$, each figure a continuous volume fraction from $2\times10^{5}$ samples per dimension, $151$ hits at $d=8$, a standard error of $6\times10^{-5}$.
Four families act as baselines throughout, the $M=3$ factor family with loadings as drawn, the same loadings after a QR orthogonalisation, the $M=5$ family, and the variable-diagonal family $\sum_\mu\xi_\mu v_\mu v_\mu^{\mathsf T}+\mathrm{diag}(\zeta)$ with $\zeta\in[0.01,0.04]^d$ at the same loadings.
Targets are drawn as $\Sigma^\star=BB^{\mathsf T}+\mathrm{diag}(\delta^\star)$ with $B_{ij}\sim\mathcal{N}(0,1/d)$ scaled by $0.35$ and $\delta^\star_i\sim U(0.01,0.04)$. The fitted family's diagonal is drawn independently from the same law and held fixed. The fitted diagonal and the loading directions are set in advance, as the register requires.
Box minima come from a corner-plus-random scan refined by local descent and cone minima from a Cholesky parameterisation. The friction sweep holds the exposure $E$ fixed by retuning $S$ at each $\rho$, at $m=5$ bits per variable, the integration step at $0.02$ or below with a floor of $3\times10^{3}$ steps.

\section{Results}
\label{sec:results}

\subsection{Infeasibility of the entry parameterisation}
\label{sec:psd}

Table~\ref{tab:psd} reports the sampled feasible fraction under the box of Sec.~\ref{sec:setup}.

\begin{table}[ht]
\caption{Fraction of the design hypercube that decodes to a positive semidefinite covariance under the entry parameterisation.}
\label{tab:psd}
\centering
\setlength{\tabcolsep}{4pt}
\begin{tabular}{lcccccc}
\toprule
$d$ & 2 & 3 & 4 & 5 & 6 & 8 \\
$M$ & 3 & 6 & 10 & 15 & 21 & 36 \\
\midrule
feasible fraction & $0.896$ & $0.635$ & $0.335$ & $0.127$ & $0.033$ & $0.0008$ \\
\bottomrule
\end{tabular}
\end{table}

On the $d=2$ register we observe $|e^{-a(k)T}|>1$ at $552$ of the $558$ infeasible points, where the unit-subnormalisation contraction construction cannot represent it and the amplitude has to be clamped.
On all $558$ the objective stays above the feasible minimum, at $0.091$ or more on the rescaled scale against $0$, so no spurious global optimum is created, a property observed here rather than guaranteed by the construction. At $d=6$ the covariance-entry register has $M=21$ variables, $105$ design qubits at $m=5$, and is not integrated here.

\subsection{The representational floor}
\label{sec:gap}

\begin{figure}[ht]
\centering
\includegraphics[width=\columnwidth]{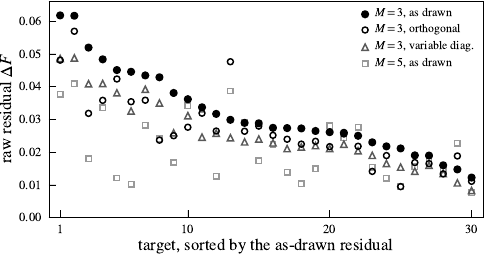}
\caption{Raw residual \eqref{eq:gap} on the thirty targets, sorted by the as-drawn value.}
\label{fig:gap30}
\end{figure}

A rank-$M$-plus-diagonal model cannot reproduce an arbitrary covariance, and the right measure of that price is the objective itself rather than a matrix norm.
Take a target $\Sigma^\star$ drawn from the full positive semidefinite cone and define the gap
\begin{equation}
\Delta F \;=\; \min_{\xi\in[0,1]^M} F\big(\Sigma(\xi)\big)
\;-\; \min_{\Sigma\succeq0} F(\Sigma),
\label{eq:gap}
\end{equation}
in units of the range of $F$ over the fitted family's own design box, the scaling used for tolerance thresholds throughout.
The second term is zero by construction, since $u^\star$ is generated from $\Sigma^\star$ by the same forward map, and a Cholesky minimisation over the cone recovers it to between $1.0\times10^{-9}$ and $7.3\times10^{-9}$ on every target.
$\Delta F$ is therefore the floor the factor family cannot go below.
Letting the diagonal vary at the same loadings, $\sum_\mu\xi_\mu v_\mu v_\mu^{\mathsf T}+\sum_i\zeta_i e_ie_i^{\mathsf T}$ with $\zeta\in[0.01,0.04]^d$, satisfies Proposition~1 equally at $m(M+d)$ qubits, and the fourth row of Table~\ref{tab:gap} measures it. The raw residual (Fig.~\ref{fig:gap30}) falls on all thirty targets, by a median factor of $0.811$.
On the family-normalised measure its median gap, $0.224$, is close to that of five fixed-diagonal factors, $0.228$. On the raw scale the five-factor family cuts deeper, median ratio $0.656$ against $0.811$.
At $d=6$ and $m=5$ the three registers cost $45$, $25$ and $105$ design qubits for variable-diagonal $M=3$, fixed-diagonal $M=5$ and covariance entries.
Fixing the diagonal at the target's own $\delta^\star$ instead gains almost nothing, so the gain of the fourth row is not diagonal matching.
The floor that remains reflects the whole dictionary, atom count, fixed directions and coefficient bounds together, not rank alone. A wider box was not tested, so the floor belongs to this box.
Table~\ref{tab:gap} reports it over thirty targets, each paired with its own factor frame.
Each configuration is normalised by its own objective range, whose medians differ by factors of $1.06$ and $1.21$ across rows, so a smaller entry does not by itself imply a smaller raw residual. We therefore also record the raw residuals, shown target by target in Fig.~\ref{fig:gap30}, and the two measures disagree on one of the sixty paired comparisons below.

\begin{table}[ht]
\caption{Normalised representation gap \eqref{eq:gap} over thirty targets at $d=6$, in units of each family's own objective range.}
\label{tab:gap}
\centering
\setlength{\tabcolsep}{6pt}
\begin{tabular}{lccc}
\toprule
 & min & median & max \\
\midrule
$M=3$, as drawn     & $0.196$ & $0.452$ & $1.095$ \\
$M=3$, orthogonal   & $0.163$ & $0.352$ & $0.744$ \\
$M=5$, as drawn     & $0.087$ & $0.228$ & $0.552$ \\
$M=3$, variable diag. & $0.089$ & $0.224$ & $0.499$ \\
\bottomrule
\end{tabular}
\end{table}

At $M=3$ the floor has median $0.452$ of the range, far above the tolerance thresholds of $0.01$ and $0.05$ that Secs.~\ref{sec:conditioning} and \ref{sec:sweep} work with. On such a target the optimiser is finding the best point of a family that cannot reach it.
On one target of the thirty the gap exceeds one, meaning the family's best point lies further from the attainable optimum than the whole variation the optimiser can see across the design box.
Paired on the same targets and on the normalised measure, five factors lower the gap in $26$ of $30$ and orthogonal directions in $28$ of $30$, by medians of $0.238$ and $0.079$. On the raw residual the counts are twenty-five and twenty-eight, the single disagreement of the sixty being a five-factor comparison.
Neither removes the floor, and the spread stays wide in every configuration.
The Frobenius picture is milder only because it fits the directions to the target, median error $0.12$ at $M=3$ against $0.79$ for a fixed random frame. The encoding is as good as the loadings supplied to it, and supplying them is a classical modelling step outside the quantum algorithm.
In our view this limitation outweighs the conditioning measured next.
The restriction is not ad hoc.
With $V=(v_1,\dots,v_M)$, \eqref{eq:factor} is $V\,\mathrm{diag}(\xi)\,V^{\mathsf T}+\mathrm{diag}(\delta)$, the factor-model covariance with loadings $V$ and independent factors \cite{fan2008high}, so the design variables are factor variances and $\xi_\mu\ge0$ is the requirement that a variance be non-negative.

\subsection{Conditioning across factor structures}
\label{sec:conditioning}

The factor form removes the infeasible region by restricting the design family, and the restricted family is ill-conditioned.
Table~\ref{tab:spread} reports thirty independent draws of the loadings at $d=6$.
The design coordinates of the target are held at $\xi^\star=(0.62,0.28,0.75)$ throughout, but $\Sigma(\xi^\star)$ and hence $u^\star$ move with the loadings, so these are thirty instances of one problem family rather than thirty encodings of one problem.
That is also why the residual vanishes at every minimiser, since each target lies in its own family by construction.

\begin{table}[ht]
\caption{Thirty independent factor structures at $d=6$, $M=3$.}
\label{tab:spread}
\centering
\setlength{\tabcolsep}{5pt}
\begin{tabular}{lccccc}
\toprule
 & min & $25\%$ & median & $75\%$ & max \\
\midrule
$\mu$      & $0.029$ & $0.084$ & $0.161$ & $0.237$ & $0.415$ \\
$\kappa$   & $3.75$  & $7.07$  & $10.12$ & $20.21$ & $79.18$ \\
$\nu=2\sqrt\mu$ & $0.343$ & $0.578$ & $0.803$ & $0.974$ & $1.289$ \\
\bottomrule
\end{tabular}
\end{table}

The local Hessian condition number at the minimiser spans a factor of $21$ and $\nu=2\sqrt\mu$ a factor of $3.75$, from nothing but a different set of factor directions at the same dimension.
We report $\nu$ rather than the exposure it implies.
Equation \eqref{eq:exposure} carries $\alpha_{\mathrm{obj}}$, but rescaling the objective to the unit interval would rescale the subnormalisation by the objective range, which differs across landscapes. Exposures are therefore comparable within a landscape and not across.

We guessed that the ill-conditioning came from non-orthogonality of the loadings and tested the guess in a paired experiment. For each of thirty draws the landscape was measured twice, once with the loadings as drawn and once with $V$ replaced by its QR factor at unchanged column norms.
The result is in Table~\ref{tab:orth}. Orthogonalisation raises $\mu$ in twenty-nine of the thirty pairs and lowers it by $0.1\%$ in the other, lifting $\sqrt\mu$ by a median factor of $1.21$, while the condition number afterwards still runs as high as $63.5$.

\begin{table}[ht]
\caption{Loadings as drawn against their QR factor, thirty paired draws at $d=6$.}
\label{tab:orth}
\centering
\setlength{\tabcolsep}{4pt}
\begin{tabular}{lccc}
\toprule
 & $\sqrt{\mu}$ ratio & $\kappa$ as drawn & $\kappa$ orthogonal \\
\midrule
minimum & $1.00$ & $3.75$  & $2.87$ \\
median  & $1.21$ & $10.12$ & $8.54$ \\
maximum & $2.47$ & $79.18$ & $63.54$ \\
\bottomrule
\end{tabular}
\end{table}

Orthogonalisation is therefore a partial remedy at best.
On the thirty out-of-family targets of Sec.~\ref{sec:gap} a separate paired test finds the QR factor lowering the raw residual in twenty-eight, by a median factor of $0.870$.
We did not measure the two effects jointly on one out-of-family optimisation.
Widening the family is the larger lever (Sec.~\ref{sec:gap}).

There is a structural reason why orthogonalisation cannot close the gap.
At the minimiser the residual vanishes to $10^{-18}$ here, so the Hessian of $\tilde F$ reduces to its Gauss--Newton part, a mode-weighted Gram matrix of the sensitivity functions $c_\mu(k)=\tfrac12(v_\mu\!\cdot\!k)^2+\tfrac{i}{2}\sum_i v_{\mu i}^2 k_i$, which depend on $v_\mu$ through a quadratic form.
Squares of orthogonal vectors are not orthogonal.
Under a rotationally symmetric mode weight,
$\mathbb{E}\big[(v_\mu^{\mathsf T}k)^2(v_\nu^{\mathsf T}k)^2\big]
= c_1\|v_\mu\|^2\|v_\nu\|^2 + c_2 (v_\mu^{\mathsf T}v_\nu)^2$,
so the first term survives even at $v_\mu\perp v_\nu$ and the Gram matrix keeps a common direction while the differences between factors stay weakly determined.
The mode weight here is not exactly isotropic, so this is an indication rather than a derivation.

\subsection{Register resolution}
\label{sec:resolution}

Table~\ref{tab:mbits} runs QHD on the seed-11 landscape, the draw closest to the median of Table~\ref{tab:spread}, against the design bit depth, with $\hat\xi$ the mode of the final measurement distribution. Entries agree between $3\times10^{3}$ and $10^{4}$ integration steps to five digits.

\begin{table}[ht]
\caption{QHD against design bit depth on the seed-11 landscape.}
\label{tab:mbits}
\centering
\setlength{\tabcolsep}{4pt}
\begin{tabular}{lcccc}
\toprule
$m$ & points & $P(\tilde F\le0.05)$ & $\|\hat\xi-\xi^\star\|$ & spacing \\
\midrule
4 & $4\,096$   & $0.9624$ & $0.0555$ & $0.0667$ \\
5 & $32\,768$  & $0.9876$ & $0.0149$ & $0.0323$ \\
6 & $262\,144$ & $0.9936$ & $0.0207$ & $0.0159$ \\
\bottomrule
\end{tabular}
\end{table}

The tolerance set $\{\tilde F\le0.05\}$ occupies $21$ to $23\%$ of the design volume. Along the weakest eigendirection it has half-width $\sqrt{2\times0.05/\mu}=0.80$ at $\mu=0.156$, almost the whole box, so the probability column says little about where the design ends up.
That probability is still rising at $m=6$, so nothing saturates there.
What does stop improving is the recovered design. The distance from $\hat\xi$ to $\xi^\star$ falls by a factor of $3.7$ from $m=4$ to $m=5$ and then worsens slightly.
Three bit depths on one landscape do not establish a floor, and finite schedule length, integration error and the choice of estimator are all untested alternatives. What the table supports is that on this landscape nothing is gained beyond $m=5$.

\subsection{The friction sweep}
\label{sec:sweep}

Curvature calibration alone does not say which multiple of $\sqrt\mu$ to take.
The classical limit \eqref{eq:damped} suggests critical damping, $\nu=2\sqrt\mu$, which maximises the decay rate of the slowest classical eigendirection.
That argument is about a classical trajectory rather than about where a wavefunction concentrates.
The equilibration bounds of \cite{catli2025efficient} in fact require friction at or below critical damping for the softest mode, $\nu\le2\sqrt\mu$, so the upper end of that range is where their analysis stops being valid. The measurement below locates the optimum inside it.
Write $\rho=\nu/(2\sqrt\mu)$ for the friction in units of critical damping.
Table~\ref{tab:sweep} sweeps $\rho$ at fixed exposure $E$ on the seed-11 landscape, the draw at $\mu=0.156$ closest to the median of Table~\ref{tab:spread}, at $m=5$ bits per variable, so every row carries the same leading objective-phase cost.
Fixing $E$ fixes neither $S$ nor $\lambda(S)$, so this compares whole schedule shapes and not damping alone.

\begin{figure}[ht]
\centering
\includegraphics[width=\columnwidth]{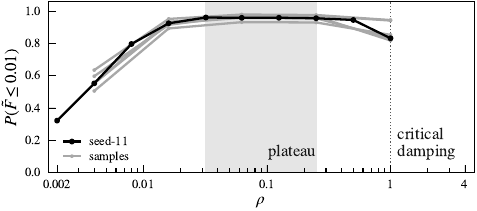}
\caption{Friction sweep at fixed objective-phase exposure. Black: ten points on the seed-11 landscape. Grey: four further draws on a coarser grid. The shaded band is the plateau $\rho\in[0.032,0.25]$.}
\label{fig:friction}
\end{figure}

\begin{table}[ht]
\caption{Friction sweep at fixed objective-phase exposure on the seed-11 landscape. Critical damping is the row $\rho=1$. At $\Delta s=0.0200$ the $3\times10^{3}$-step floor is inactive.}
\label{tab:sweep}
\centering
\setlength{\tabcolsep}{2pt}\footnotesize
\begin{tabular}{lcccccc}
\toprule
$\rho$ & $S$ & $\lambda(S)$ & $\Delta s$ & $P(\tilde F\le0.01)$ & $P(\tilde F\le0.05)$ & $\mathbb{E}[\tilde F]$ \\
\midrule
$0.004$ & $1424$ & $90$      & $0.0200$ & $0.552$ & $0.947$ & $0.0152$ \\
$0.016$ & $465$  & $357$     & $0.0200$ & $0.925$ & $0.998$ & $0.0038$ \\
$0.032$ & $260$  & $712$     & $0.0200$ & $\mathbf{0.961}$ & $0.998$ & $0.0023$ \\
$0.125$ & $80$   & $2779$    & $0.0200$ & $0.959$ & $0.997$ & $\mathbf{0.0018}$ \\
$0.25$  & $44$   & $5556$    & $0.0146$ & $0.956$ & $0.996$ & $0.0019$ \\
$1$     & $13$   & $22\,222$ & $0.0042$ & $0.832$ & $0.988$ & $0.0061$ \\
$4$     & $3.6$  & $88\,886$ & $0.0012$ & $0.151$ & $0.747$ & $0.0387$ \\
\bottomrule
\end{tabular}
\end{table}

Across the five draws the best tested friction lies between $0.032$ and $0.25$ of critical damping (Fig.~\ref{fig:friction}), the seed-11 dense sweep peaking at $\rho=0.032$ with $0.016<\rho^\star<0.0625$ and $\mathbb{E}[\tilde F]$ smallest at $\rho=0.125$, and across that interval the surface is flat, $P(\tilde F\le0.01)$ varying by $0.005$ and $\mathbb{E}[\tilde F]$ by $6\times10^{-4}$.
Below the plateau the collapse is fast, the total variation to the uniform falling from $0.95$ to $0.67$ at $\rho=0.002$ as lower $\nu$ at fixed $E$ lowers $\lambda(S)$ and the schedule stops annealing.
On the coarser grid $\rho\in\{0.004,0.016,0.0625,0.25,1\}$ the other four draws peak at $\rho=0.0625$ three times and at $0.25$ once, with $\mathbb{E}[\tilde F]$ minimal at $0.25$ on all four, too coarse to resolve the seed-11 peak.
Against critical damping the plateau is worth $3$ to $15$ points of success probability and $2.3$ to $4.8$ times in expected objective.

Over all five landscapes, seven frictions $\rho\geq0.25$ and three statistics, the largest change from $3\times10^{3}$ to $10^{4}$ integration steps is $1.9\times10^{-4}$, and refining $\Delta s$ to $0.0067$ on the plateau moves $P$ by at most $2.1\times10^{-4}$ and $\mathbb{E}[\tilde F]$ by $8.1\times10^{-6}$, leaving the peak at $\rho=0.032$.
The distributions for $\tilde F$ and $\tilde F+b$ agree pointwise to $1.3\times10^{-14}$ at $b=1$ and $1.2\times10^{-11}$ at $b=10^{3}$.
With $\nu=2\sqrt{\hat\mu}$ the abscissa of Table~\ref{tab:sweep} is $\sqrt{\hat\mu/\mu}$ at fixed $E$, so the table prices a wrong curvature estimate, $\hat\mu/\mu=1/16$ still yielding $P(\tilde F\le0.05)=0.996$ while $16$ yields $0.747$, a $24\%$ loss.
Under the fixed-exposure retuning used here, the simulations are therefore substantially more tolerant of underestimating the curvature than of overestimating it.

\section{Discussion}
\label{sec:disc}
 
\noindent\textbf{What fixed exposure fixes.}
One quantity is held fixed by construction, and two are not, since $E$ fixes the leading objective-phase term only. $S$ and $\lambda(S)$ move with $\rho$, as Table~\ref{tab:sweep} shows, and the kinetic simulation cost grows with $S$, so equal exposure is not equal end-to-end cost.
 
\noindent\textbf{Operator dependence.}
A periodic spectral Laplacian makes the design box a torus, and under it $P(\tilde F\le0.01)$ peaks near $\rho=0.5$ to $1$ and exceeds $0.999$ where the interval operator gives $0.95$. Under the operator of Sec.~\ref{sec:protocol} the peak moves down by more than a decade, to the plateau of Sec.~\ref{sec:sweep}.

\noindent\textbf{Scope.}
These are five instances of one benchmark under one stopping rule.

\noindent\textbf{Classical cost of the calibration.}
Obtaining $\mu$ requires $\xi^\star$, $F_{\min}$ and $F_{\max}$, a classical solution of the design problem, so the calibration is a small-$M$ diagnostic, not a general recipe.
 
\noindent\textbf{Stability under spatial refinement.}
Refining a $d=4$ landscape through $N_x=4,8,16,32$ leaves $\mu$ between $0.0149$ and $0.0150$, the other two Hessian eigenvalues settling at $(0.494,\,1.486)$ from $N_x=8$, and two $d=6$ resolutions give $\mu=0.1560$ and $0.1557$, which is what admits the coarse grids.
 
\noindent\textbf{Dimension dependence.}
The curvature is not monotone in $d$, $0.015$, $0.156$, $0.095$ and $0.076$ at $d\in\{4,6,8,10\}$, and the spread across structures at fixed $d$ is larger still, so nothing supports extrapolation in $d$.

\section{Conclusion}

The oracle construction for coherent design pipelines assumes affine generators with positive semidefinite Hermitian parts, which covariance entries fail off the diagonal, while an affine combination of positive semidefinite atoms supplies both at every register point, factor form its rank-one case.
For the tested rank-three dictionary the out-of-family floor has median $45\%$ of the family's own range, falling to $22\%$ with a variable diagonal, and it outweighs the conditioning spread.
 
The additive gauge symmetry is exact, the mean rule breaks it, and the curvature rule restoring it leaves the constant in $\nu=c\sqrt\mu$ open. At fixed exposure the best tested friction lies four to thirty-one times below critical damping on all five draws, the dense draw showing a broad plateau.
The recommendation is to fix the additive freedom and, given a curvature estimate, to prefer underestimation under this retuning, a five-draw diagnostic rather than a calibration.

\section*{Acknowledgement}

Generative AI (Claude, Anthropic~\cite{anthropic2026claude}) assisted in drafting and editing all sections and in preparing the public code. The author verified all content and takes full responsibility for it.

\noindent\textbf{Data and code.}
Scripts and stored run logs for every table and figure: \url{https://github.com/khoshisashi/qhd-curvature-friction}.
No quantum hardware and no circuit simulation enter anywhere.

\newpage
\bibliographystyle{IEEEtran}
\bibliography{references}

\end{document}